\documentclass[
    ,final            
  ]
  {aipproc}

\layoutstyle{6x9}

\newcommand{\beq}{\begin{eqnarray}}
\newcommand{\eeq}{\end{eqnarray}}

\newcommand{\ie}{{\it i.e.\ }}

\newcommand{\pbp}{\langle \bar{\psi} \psi \rangle}

\begin{document}

\title{Order, Disorder and Confinement}

\classification{11.15.Ha, 12.38.Aw, 14.80.Hv, 64.60.Cn}
\keywords      {QCD, Confinement, Deconfinement Transition}

\author{M. D'Elia}{
  address={Dipartimento di Fisica dell'Universit\`a di Genova and INFN,
Sezione di Genova, Via Dodecaneso 33, I-16146 Genova, Italy }
}

\author{A. Di Giacomo\thanks{Talk presented by A. Di Giacomo}}{
  address={ Dipartimento di Fisica dell'Universit\`a di Pisa and INFN, 
Sezione di Pisa, largo Pontecorvo 3, I-56127 Pisa, Italy}
}

\author{C. Pica}{
  address={ Dipartimento di Fisica dell'Universit\`a di Pisa and INFN, 
Sezione di Pisa, largo Pontecorvo 3, I-56127 Pisa, Italy}
}

\begin{abstract}
Studying the order of the chiral transition for $N_f=2$ is of
fundamental
importance to understand the mechanism of
color confinement.  We present results of a numerical
investigation on the order
of the transition by use of a novel strategy in finite size
scaling analysis.  
The specific heat and a number of susceptibilities
are compared with the possible critical behaviours. 
A second order transition
in the $O(4)$ and $O(2)$ universality classes are
excluded. Substantial evidence emerges for a first order transition.
Results are in agreement with those found by studying the scaling 
properties of a disorder parameter related to the dual 
superconductivity mechanism of color confinement.
\end{abstract}

\maketitle


\section{Introduction}

\begin{figure}
  \includegraphics[height=.3\textheight]{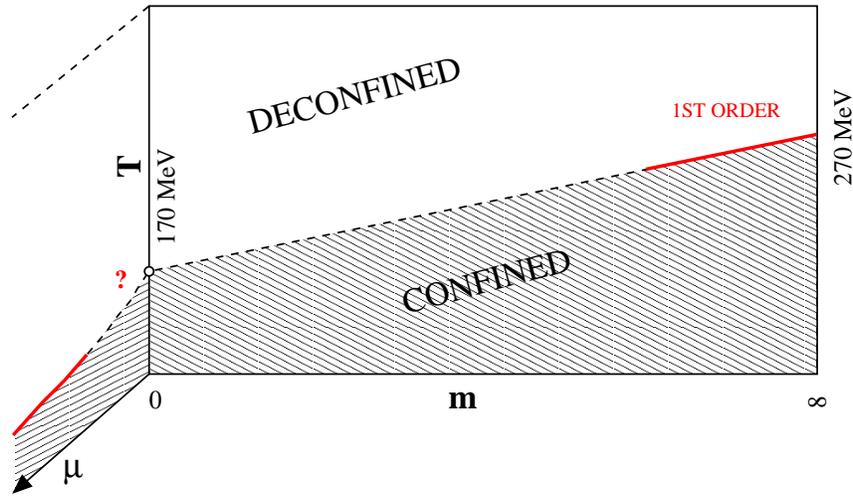}
\caption{Schematic phase diagram of $N_f=2$ QCD.}\label{PHDIA}
\end{figure}

Experiments indicate that confinement is an
absolute property of matter. Indeed the upper limit on the number
of free quarks per proton is $R = n_q/n_p \leq 10^{-27}$, while
$R \sim 10^{-12}$ is expected from the Standard Cosmological Model.  
A reduction factor $10^{-15}$ is difficult to explain in natural
ways unless $R = 0$, which means that confinement is 
related to some symmetry of the QCD vacuum. This also implies
that the deconfining transition is associated to a change of symmetry
of the vacuum, \ie it is an order-disorder transition.
This scenario must be confronted with direct studies of the
deconfining phase transition: experimental studies going on 
with heavy ion collisions have not given definite answers yet 
and most of our knowledge relies on numerical simulations of
QCD on the lattice. From this point of view, the case of two
light degenerate dynamical flavours is of special
interest. A schematic view of the phase diagram for $N_f = 2$ 
is shown in Fig.~\ref{PHDIA}: 
$m$ is the quark mass and  $\mu$ is the baryon chemical potential.

In the $\mu = 0$ plane, as $m \to \infty$ quarks decouple and the
system tends to the quenched limit. There the deconfining transition
is an order-disorder first order
phase transition, $Z_3$ is an associated
symmetry and the Polyakov line
$\langle L \rangle$ is an order parameter. 
$Z_3$ is explicitely broken by the inclusion of dynamical quarks 
and $\langle L \rangle$ is not a good
order parameter, even if it works as such
for quarks masses down to $m \simeq 2.5 - 3$ GeV.

At $m \simeq 0$ there is chiral phase transition at $T_c\simeq
170$ MeV, from a low temperature phase where chiral symmetry is
spontaneously broken to an high temperature phase in which it is
restored: 
the corresponding order parameter is the chiral
condensate $\pbp$.  At some
temperature $T_A\geq T_c$ also the $U_A(1)$ symmetry, broken
by the anomaly, is expected to be effectively restored.
It is not clear which relation exists between the chiral transition and the
deconfining transition: empirically the Polyakov line has a rapid
increase at the transition temperature, indicating deconfinement.
The transition line in Fig.~\ref{PHDIA} is defined by
the maxima of a number of susceptibilities ($C_V$, $\chi_{m}$,
$\dots$), all coinciding within errors, which indicate a rapid
variation of the corresponding parameters across the line.

At $m \simeq 0$ a renormalization group analysis plus
$\epsilon$-expansion techniques can be made, assuming that the 
relevant degrees of
freedom for the chiral transition are scalar and pseudoscalar
fields~\cite{wilcz1,wilcz2,wilcz3}. If the  $U_A(1)$ symmetry is 
effectively restored, 
\ie if the $\eta '$ mass vanishes at $T_c$, then there is no IR stable
fixed point and the phase transition is first order; if 
not an IR
fixed point exists, which can produce a second order
phase transition in the $O(4)$ universality class.

In the first case the transition is first order also at $m \neq
0$ and most likely up to $m = \infty$.
In the second case a phase transition is only present at $m = 0$,
which goes into a continuous crossover as $m \neq 0$: that means that
one can move continuously from confined to deconfined and that 
no true order parameter exists. This would be in contradiction with
the deconfinement transition being associated to a 
change of symmetry: the issue is therefore fundamental.

The problem has been investigated on the lattice by several groups with
staggered~\cite{fuku1,fuku2,colombia,karsch1,karsch2,jlqcd,milc} or
Wilson~\cite{cp-pacs} fermions. The strategy used has either been to
search for signs of discontinuity at the transition, or to study the
scaling with $m$ of different susceptibilities, or to
study the magnetic equation of state. No clear discontinuities
have been observed, but also no conclusive agreement of scaling with
$O(4)$ critical indexes. 
We present the results of a big numerical effort aimed at 
clarifying the issue. We use non improved Kogut--Susskind action and lattices $4 \times
L_s^3$ with $L_s = 16,20,24,32$. A full account of our results can be
found
in~\cite{nostrolavoro}.

\section{Results}

The theoretical tool to investigate the order of a phase transition is
finite size scaling. The extrapolation from
finite size $L_s$ to the thermodynamical limit is
governed by critical indexes, which identify the order and the
universality class of the transition.
Around the chiral transition the system has two fundamental lengths:
the correlation length $\xi$ and the inverse quark mass
$1/m_q$. $\xi$ is usually traded with the reduced temperature 
$\tau = 1 - T/T_c$,  $ \xi \simeq\tau^{-\nu} $ as $\tau \to 0$.
The effective action
depends on the order parameter, as dictated by the symmetry, and as
$\tau \to 0$ irrelevant terms can be neglected; the correlators
of the order parameter describe  the thermodynamics.
The most important quantity is the specific heat, which shows 
the correct critical behaviour independently
of the identification of the order parameter.

For the specific heat and the susceptibility of the order
parameter the scaling laws are 
\beq
C_V - C_0 \simeq  L_s^{\alpha/\nu} \phi_c \left(\tau L_s^{1/\nu}, am_q L_s^{y_h} \right) \, ;
\label{scalcal} \\
\chi \simeq L_s^{\gamma/\nu} \phi_\chi \left(\tau L_s^{1/\nu}, am_q L_s^{y_h} \right) \, .
\label{scalord}
\eeq
$C_0$ stems from an additive renormalization. 
An alternative way to write them is
\begin{eqnarray}
C_V - C_0 \simeq  L_s^{\alpha/\nu} \tilde\phi_c \left(\tau (am_q)^{-1/(\nu y_h)},am_q L_s^{y_h} \right)\label{scalcal2A} \\
\chi \simeq L_s^{\gamma/\nu} \tilde\phi_\chi \left(\tau (am_q)^{-1/(\nu y_h)}, am_q L_s^{y_h} \right) \, .
\label{scalord2A}
\end{eqnarray}

\begin{figure}[t!]
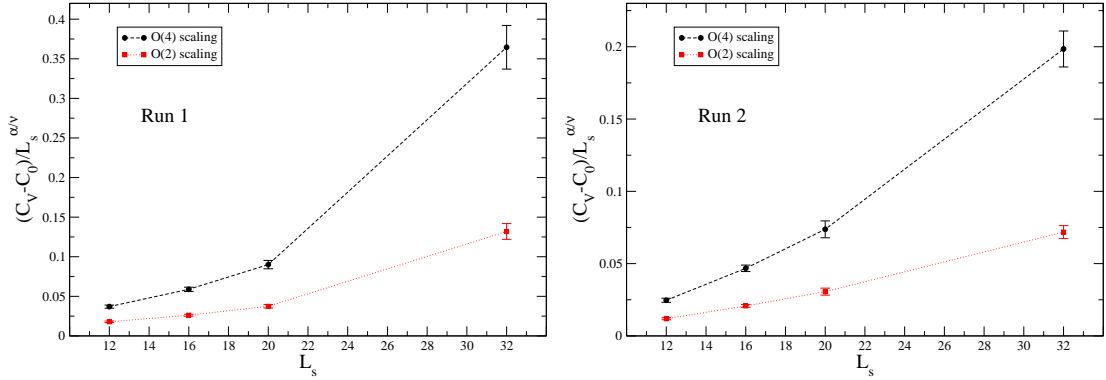

\includegraphics[height=5.0cm,width=7.2cm]{Cv_max_Run1.eps}\hspace{0.1cm}
\includegraphics[height=5.0cm,width=7.2cm]{Cv_max_Run2.eps}
\caption{Specific heat peak value for Run1
(left) and for Run2 (right), divided by the appropriate powers of
$L_s$ to give a constant. Both the $O(4)$
and $O(2)$ critical behaviors are displayed.}\label{R12}
\end{figure}

The values of the indexes characterize the transition: the values
relevant to our analysis are listed in
Table~\ref{CRITEXP}. $O(4)$ is the symmetry expected if the 
transition is second order, but it can break down to $O(2)$ by 
lattice discretization for Kogut--Susskind fermions~\cite{karsch1} at
non zero lattice spacing.

\begin{table}[b!]
\begin{tabular}{cccccc}
\hline & {\bf $y_h$} & {\bf $\nu$} & {\bf $\alpha$} & {\bf $\gamma$} &
{\bf $\delta$} \\
\hline $O(4)$ & 2.487(3) & 0.748(14) & -0.24(6) & 1.479(94) & 4.852(24)\\
$O(2)$ & 2.485(3) & 0.668(9) & -0.005(7) & 1.317(38) & 4.826(12)\\
$MF$ & $9/4$ & $2/3$ & 0 & 1 & 3\\
$1^{st} Order$ & 3 & $1/3$ & 1 & 1 & $\infty$\\
\hline
\end{tabular}
\caption{Critical exponents.}\label{CRITEXP}
\end{table}


\begin{figure}[t!]
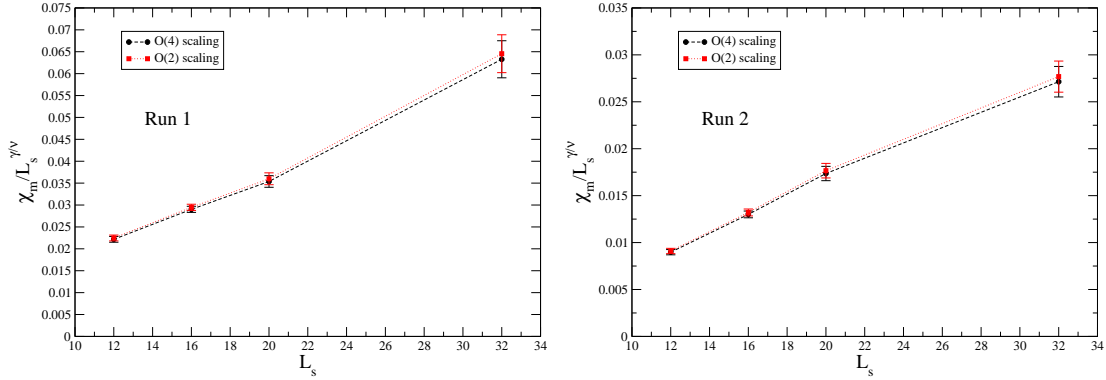

\includegraphics[height=5.0cm,width=7.2cm]{Chi_max_Run1.eps}\hspace{0.1cm}
\includegraphics[height=5.0cm,width=7.2cm]{Chi_max_Run2.eps}
\caption{The same as figure \ref{R12} for the chiral susceptibility
$\chi_m$.}\label{R12bis}
\end{figure}

The scaling analysis is made difficult by the presence of two
independent scales. The attitude taken in the previous
literature has been to assume the volume large enough so to neglect
the dependence on $L_s$: since at fixed $am_q$, $\beta$
the susceptibilities must be analytic in the thermodynamical limit, 
at large $L_s$ the dependence on $am_q L_s^{y_h}$ must cancel the
dependence on $L_s$ in front of the scaling functions in
Eq.s (\ref{scalcal}) and (\ref{scalord}). It follows that
\begin{eqnarray}
C_V - C_0 \simeq  (am_q)^{-\alpha/(\nu y_h)} f_c \left(\tau (am_q)^{-1/(\nu y_h)}\right) \label{scalcal1} \\
\chi \simeq  (am_q)^{-\gamma/(\nu y_h)} f_\chi \left(\tau (am_q)^{-1/(\nu y_h)}\right) \, .\label{scalord1}
\end{eqnarray}
The peaks of $(C_V - C_0)$ and of $\chi$ should then scale as
\beq
(C_V - C_0)_{\rm max} \propto (am_q)^{-\alpha/(\nu y_h)} \,; \,\,\,\,\,\,\,\,\,\,
\chi_{\rm max} \propto (am_q)^{-\gamma/(\nu y_h)}
\label{scalmax1}
\eeq
as $am_q\rightarrow 0$. The positions of the maxima scale as 
$
\tau (am_q)^{-1/(\nu y_h)} = {\rm const} \label{pcscale} \, .
$

One can also consider to keep $\tau L_s^{1/\nu}$ fixed 
while taking
$a L_s \gg 1/m_\pi$. This assumption should work better if $L_s$ is
still comparable to the correlation length, which may be the case
close enough to the critical point. In this case the scaling laws are
\begin{eqnarray}
C_V - C_0 \simeq  (am_q)^{-\alpha/(\nu y_h)} f_c \left(\tau  L_s^{1/\nu} \right)
\label{scalcal2} \,; \,\,\,\,\,\,\,\,\,
\chi \simeq  (am_q)^{-\gamma/(\nu y_h)} f_\chi \left(\tau  L_s^{1/\nu} \right) \, .
\label{scalord2}
\end{eqnarray}
Eq.s~(\ref{scalmax1}) stay unchanged, the positions
of the maxima scale as 
$
\tau L_s^{1/\nu} = {\rm const} \label{pcscale1} \, 
$
and the width of the peaks are volume dependent.

We have instead followed a novel strategy which does not rely on any
assumption: we have performed a number of simulations at
different values of $L_s$ and $a m_q$ keeping the variable
$a m_q L_s^{y_h}$ fixed and studying the scaling with the respect to 
the other one. In doing so one has to assume a value
for $y_h$: we have chosen that expected for $O(4)$, which is 
the same within errors as for $O(2)$.
From Eq.s (\ref{scalcal}) and (\ref{scalord}) it follows that, at fixed
$am_q L_s^{y_h}$
\beq
(C_V - C_0)_{\rm max} \propto L_s^{\alpha/\nu} \,; \,\,\,\,\,\,\,\,\,\,
\chi_{\rm max} \propto L_s^{\gamma/\nu} \, .\label{scalmax}
\eeq
as $L_s \rightarrow \infty$.
If $O(4)$ or $O(2)$ is the correct symmetry, the values of
$\alpha/\nu$ and $\gamma/\nu$ should be consistent with the
corresponding values listed in Table~\ref{CRITEXP}.
We have run two such sets of Monte
Carlo simulations, called in the
following Run1 and Run2, with $am_q L_s^{y_h}=74.7$
and $am_q L_s^{y_h}=149.4$ respectively. The spatial lattice sizes
$L_s$ used for each of the two sets are $L_s=12, 16, 20, 32$, the 
standard hybrid R algorithm~\cite{HybridR} has been used to update 
configurations.  
In Figs.~\ref{R12} and~\ref{R12bis} we show the peak values of the specific
heat and of the chiral susceptibility, divided by the appropriate power of 
$L_s$, as a function of $L_s$ (see
Eq.~\ref{scalmax}): scaling is clearly violated, $O(4)$ and
$O(2)$ universality classes are excluded.

\begin{figure}[t!]
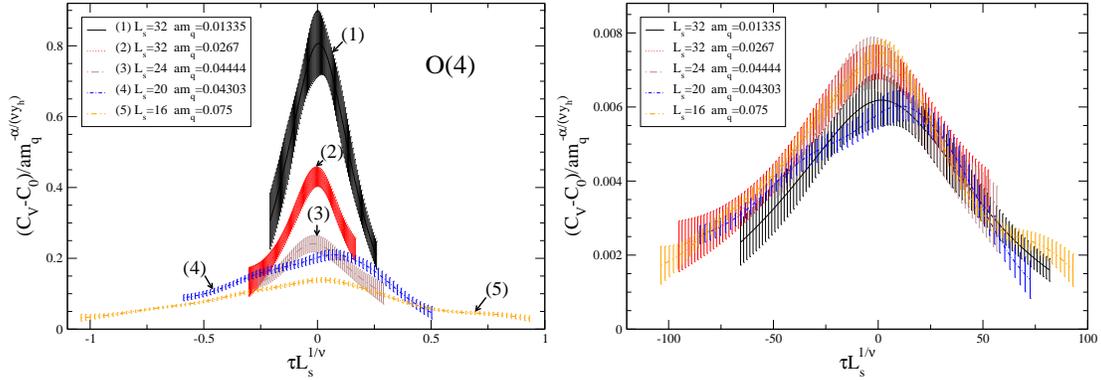

\includegraphics[height=5.0cm,width=7.2cm]{Cv_O4-Ls.eps} 
\includegraphics[height=5.0cm,width=7.2cm]{Cv_1st.eps}
\caption{Comparison of specific heat scaling,  Eq.\ref{scalcal2}, for 
$O(4)$ (left) and first order (right).}\label{fullscal}
\end{figure}

An alternative way to study the order of a transition is to look 
at scaling of pseudocritical couplings: one can try the two
alternative scaling laws $\tau_c = k_\tau L_s^{1/\nu}$ or 
$\tau_c = k_\tau' (a m_q)^{1/(\nu y_h)}$ .
 The physical temperature
$T = 1/ ( L_t a(\beta,m_q))$ is a function of both $\beta$ 
and $a m_q$, so that the reduced temperature $\tau$ can be expanded
as a power series in $(\beta - \beta_0)$ and in $a m_q$, where
$\beta_0$ is the chiral critical coupling. Only the
linear term in $\beta$ was considered in the previous literature. We
have found that the following terms are sufficient to fit the data  
\beq
\tau \propto (\beta_0 - \beta )+ k_m am_q + k_{m^2} (am_q)^2 + k_{m\beta} am_q (\beta_0 - \beta ) \, .
\label{taudef2}
\eeq
Our result is that it is not possible, within the present mass range, to
discriminate among the possible critical behaviours by looking at
pseudocritical couplings only. The
inclusion of other terms in Eq.~(\ref{taudef2}), besides the one 
linear in $\beta$ solely considered in previous literature, 
is however crucial to obtain any scaling at all.

We can also use our data to perform the scaling analysis in the same
way
as done in previous literature, \ie by making some assumption on 
the reaching of the thermodynamical limit: no universality class 
is chosen a priori in that case and one can test all the 
possible critical behaviours. We have found that assuming that 
$a L_s \gg 1/m_\pi$ but still $L_s \sim \xi$, \ie 
Eqs.~(\ref{scalcal2}), works better: this
is reasonable when $\xi$ goes large. We have added to the data
from Run1 and Run2, those from two other simulations performed at 
$L_s = 16$, $a m_q = 0.01335$ and $L_s = 24$, $a m_q = 0.04444$.
In Fig.\ref{fullscal} we show the scaling obtained for the specific
heat: $O(4)$ is again clearly excluded, while a good agreement is found 
with a weak first order critical behaviour. A similar behaviour is
observed for the chiral susceptibility.

\begin{figure}[t!]
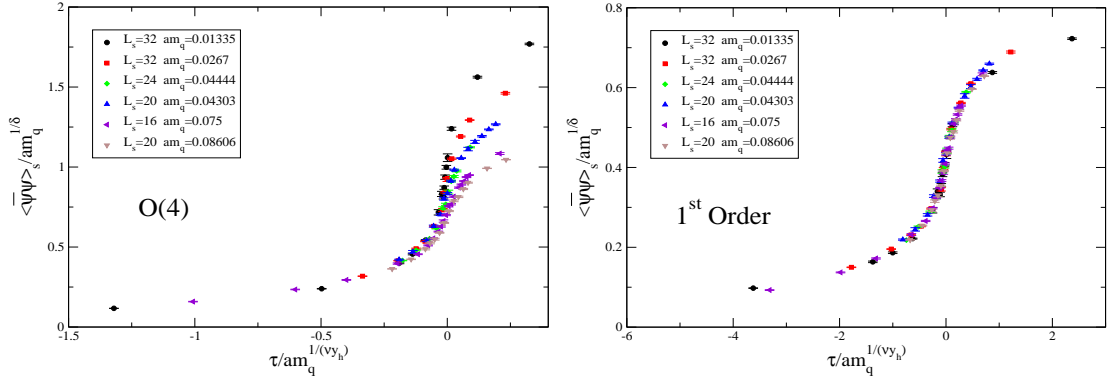

\includegraphics[height=5.0cm,width=7.2cm]{eqstO4.eps} \
\includegraphics[height=5.0cm,width=7.2cm]{eqst1st.eps}
\caption{Scaling of the equation of state for 
$O(4)$ (left) and first order (right).}\label{eqstate}
\end{figure}

We have also studied the magnetic equation of state, \ie the scaling of
$\pbp - \pbp_0 = a m_q^{1/\delta} F(\tau m_q^{1/\nu y_h})$. Results are 
shown in Fig.~\ref{eqstate}: again the first order behaviour describes
well the data while the second order is excluded.

\begin{figure}[t!]
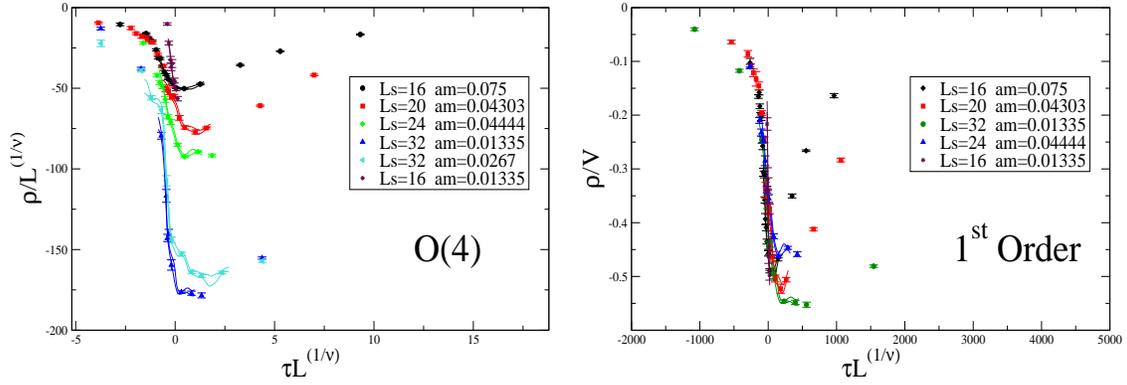

\includegraphics[height=5.0cm,width=7.2cm]{Rho_O4.eps} \hspace{0.2cm}
\includegraphics[height=5.0cm,width=7.2cm]{Rho_1st.eps}
\caption{Scaling of the susceptibility $\rho$ of the disorder 
parameter $\langle \mu \rangle$ for 
$O(4)$ (left) and first order (right).}\label{monofig}
\end{figure}

Even if we have still not found 
clear signs of discontinuities in physical observables, 
our results give some evidence for a weak first order chiral
transition which would persist also at $m \neq 0$. That would 
be in agreement with the deconfining transition being order-disorder
and with confinement being associated to some symmetry of the QCD
vacuum. This naturally leads to look for such a symmetry and for an 
order parameter which describes the deconfining transition both
with and without dynamical quarks. One candidate is dual
superconductivity
of the vacuum and the related parameter is the vacuum expectation 
value of a magnetically charged operator
$\langle \mu \rangle$.
Indeed it has been shown that it is good order parameter also in full 
QCD~\cite{fullmono1} and its critical behaviour across the
$N_f = 2$ transition has been studied in Ref.~\cite{fullmono2}. 
In Fig.~\ref{monofig} we show the scaling of its susceptibility 
$\rho = d/d\beta \ln \langle \mu \rangle$: again results seem
to indicate a first order critical behaviour.

\section{Conclusions}
We have argued that the study of the order of the chiral phase
transition for $N_f = 2$ is of fundamental importance to understand
confinement. We have shown that the analysis of pseudocritical 
coupling alone cannot discern between the possible critical
behaviours. By adopting a novel strategy which reduces the finite
size scaling analysis to a one scale problem, we have been
able to exclude a $O(4)$ ($O(2)$) second order critical behaviour,
while we have found consistency with a weak first order critical
behaviour both in the scaling of susceptibilities and in the equation
of state. This would be in agreement with confinement being an 
absolute property of matter related to some symmetry and with the
deconfining transition being order-disorder:
indeed the analysis of the scaling properties of the susceptibility of
an 
order parameter associated with the dual superconductivity mechanism
of color confinement leads to analogous results.  
 However we have  
still not found any clear evidence for discontinuities in physical
observables. The issue is still open and we plan in the future 
to investigate it more deeply 
by making simulations with improved actions and
algorithms and with $a m_q L_s^{y_h}$ fixed according to 
first order.

\begin{theacknowledgments}
This work has been partially supported by MIUR, Program ``Frontier
problems in the Theory of Fundamental Interactions''. 
\end{theacknowledgments}

\end{document}